\documentclass[twocolumn,showpacs,preprintnumbers,amsmath,amssymb]{revtex4}

\newcommand{\be}{\begin{equation}}
\newcommand{\ee}{\end{equation}}
\newcommand{\bea}{\begin{eqnarray}}
\newcommand{\eea}{\end{eqnarray}}
\newcommand{\ba}{\begin{array}}
\newcommand{\ea}{\end{array}}

\def\bbox{{\,\lower0.9pt\vbox{\hrule \hbox{\vrule height 0.2 cm
\hskip 0.2 cm \vrule height 0.2 cm}\hrule}\,}}
\newcommand{\dsl}{\pa \kern-0.5em /}

\usepackage{graphicx}
\usepackage{dcolumn}
\usepackage{bm}

\begin{document}

\preprint{UG-09-01, DAMTP-2008-115, arXiv:0901.1766}

\title{Massive Gravity in Three Dimensions}

\author{Eric A.~Bergshoeff and Olaf Hohm}
\affiliation{%
Centre for Theoretical Physics, University of Groningen,\\
Nijenborgh 4, 9747 AG Groningen, The Netherlands.
}%

\author{ Paul K.~Townsend}
\affiliation{ Department of Applied Mathematics and
Theoretical Physics,\\
Centre for Mathematical Sciences, University of Cambridge,\\
Wilberforce Road, Cambridge, CB3 0WA, U.K.
}%


\begin{abstract}

A particular higher-derivative extension of the Einstein-Hilbert
action in three spacetime dimensions is shown to be equivalent at
the linearized level to the (unitary) Pauli-Fierz action for a
massive spin-2 field.  A more general model, which also includes
`topologically-massive' gravity as a special case, propagates the two spin 2 
helicity states with different masses.  We discuss the extension to massive
${\cal N}$-extended supergravity, and we present a `cosmological' extension
that admits an anti-de Sitter vacuum.

\end{abstract}

\maketitle

For some purposes it is useful to think of Einstein's theory of
gravity, General Relativity (GR), as a model for the consistent
interaction of a massless spin 2 field on a four-dimensional ($4D$) Minkowski 
`background'. This perspective makes it clear that  the
quanta associated with gravitational waves, i.e. gravitons, are
massless particles with two independent polarization states, of
helicity $\pm2$. Unfortunately, the quantum theory of gravitons is
non-renormalizable, a fact that has led many theorists to consider GR
and its variants in three dimensions ($3D$) because one expects
less severe short-distance behaviour in a lower dimension. Pure GR is
perhaps too simple for this purpose because its linearization on a
Minkowski vacuum yields an equation that propagates no physical
helicity states \cite{Deser:1983tn}. A popular modification of GR in
$3D$ is the `topologically-massive gravity'  (TMG), which  complements the 
Einstein-Hilbert (EH) action with a Lorentz Chern-Simons (LCS) term \cite{Deser:1981wh}, 
thus breaking parity as well as introducing a new mass scale. Linearization  yields a
third-order  wave-equation but, remarkably, the theory is 
unitary and propagates a {\it single} massive mode of helicity
$\pm2$, the sign depending on the sign of the LCS
term. 

The main aim of this paper is to  present a different, {\it parity-preserving}, 
variant of $3D$ GR  that describes, on quantization, a unitary interacting theory 
of gravitons, each of which has {\it two} polarization states, of helicity
$\pm2$, as in $4D$ GR except that the $3D$ graviton is massive. This
is so despite the fact that the field equations are fourth-order in derivatives, 
because linearization of our  `massive gravity'  theory yields a free
 `fourth-order'  theory with precisely the required physical content; we confirm
 this result by a simple proof of its equivalence to  the ($3D$) Pauli-Fierz (PF)  
 `second-order' theory for a  massive  spin 2 field. In $4D$, this type of `higher-derivative' 
 theory  is not unitary but it {\it is} renormalizable \cite{Stelle:1976gc}, and this implies
 super-renormalizability in $3D$.
 
 The representation theory of the Poincar\'e group is essentially the same for massive 
 $3D$ particles as it is for massless $4D$ particles. However, the CPT theorem in $4D$
 implies that every state of helicity $h$ is accompanied by a state of helicity $-h$, with the 
 same mass. In contrast, the masses may differ in $3D$, at the cost of violating parity. 
Furthermore, given a pure spin $s$ theory, with helicities $\pm s$, we may take the mass of one 
helicity state to infinity, thereby arriving at a theory describing a single helicity $s$ state. 
In the $s=1$ case, this decoupling of $3D$ helicity states is reflected in the fact that
the Proca equation factorizes into two  first-order equations, each 
describing one helicity  state \cite{Townsend:1983xs};  the helicity
content of this `square-root'  of the Proca theory is therefore identical to that of the topologically 
massive spin 1 theory \cite{Deser:1982vy}, to which it may be shown to be equivalent via a 
`master action'  \cite{Deser:1984kw}.  A similar factorization occurs for the 
$3D$ Pauli-Fierz equation for spin 2 \cite{Aragone:1986hm},  and the resulting first-order equation has been 
shown \cite{Dalmazi:2008zh}  to be equivalent to  linearized TMG via an equivalence of both to an intermediate `self-dual'   theory \cite{Deser:1990ay}.  In other words, the linearized TMG field  equations are equivalent   to the `square-root'  of  field equations that are themselves equivalent  to the linearized equations of  our new `massive gravity' theory.  Here we unify the spin 2 equivalences that underlie this interpretation by means of  a `triple-master' action. 

We begin with a presentation of the new massive gravity (NMG) theory. Let $g_{\mu\nu}$ ($\mu,\nu=0,1,2$) be the $3D$ spacetime metric, with determinant $g$, and let $R_{\mu\nu}$ be its  Ricci curvature tensor, which determines not only the Ricci scalar $R= g^{\mu\nu}R_{\mu\nu}$ but also the full Riemann tensor. We choose the $(+--)$ metric signature.   Now consider the action
 \be\label{ouraction}
 S = \frac{1}{\kappa^{2}}\int \! d^3 x\, \sqrt{g} \left[ R  +  \frac{1}{m^{2}} K \right]\, , 
 \ee
 where
 \be\label{defK}
 K =  R_{\mu\nu}R^{\mu\nu} - \frac{3}{8}R^2\, . 
\ee
The constant $\kappa$, which has mass dimension $[\kappa]=-1/2$  in
fundamental units,  is the $3D$ analog of the square root of
Newton's constant, while $m$ is a `relative' mass parameter, which
could be traded for the effective dimensionless coupling constant
$m\kappa^2$, as for TMG  \cite{Deser:1990bj}.  Also in common with that theory is the `wrong' 
sign for the EH term.  Remarkably, this `higher-derivative' theory 
is unitary, with  field quanta that are massive spin 2 particles, each with two polarization
states, of helicity $2$ and $-2$.  

This result  may be established in various ways. Let us begin by considering  the field equations. These are
\be\label{nonlin}
2m^2 G_{\mu\nu} + K_{\mu\nu} =0\, , 
\ee
where $G_{\mu\nu} = R_{\mu\nu} - \frac{1}{2}g_{\mu\nu} R$ is the Einstein tensor, and
\begin{eqnarray}\label{Kcov}
  K_{\mu\nu}  &=& 2D^2
  R_{\mu\nu}-\frac{1}{2}\left(D_{\mu}D_{\nu}R+g_{\mu\nu} D^2 R\right) 
   - 8R_{\mu}{}^{\rho}R_{\nu\rho}  \nonumber\\
  &&  + \frac{9}{2} R R_{\mu\nu}
  +  g_{\mu\nu}\left[ 3 R^{\rho\sigma}R_{\rho\sigma} 
  - \frac{13}{8}R^2 \right]\, , 
 \end{eqnarray}
 where $D_\mu$ is the usual Levi-Civita covariant derivative, and $D^2 \equiv D^\mu D_\mu$. 
 As a consequence of the diffeomorphism invariance of the action, we have the Bianchi-type identity
 $D^\mu K_{\mu\nu} \equiv 0$. A {\it special} feature of the scalar $K$, and the tensor $K_{\mu\nu}$ derived from it, is that 
\be\label{traceK}
g^{\mu\nu} K_{\mu\nu} = K\, . 
\ee
As a consequence, the trace of (\ref{nonlin}) yields
\be\label{trace}
m^2R =K\, . 
\ee
Note, in particular, the absence on the right-hand side of a $D^2R$ term,  which would contribute to the linearized equation if it were present. 

The next step is to linearize the field equations about the  Minkowski vacuum solution that they obviously admit 
by writing $g_{\mu\nu} = \eta_{\mu\nu} + \kappa h_{\mu\nu}$ for  Minkowski metric $\eta_{\mu\nu}$ and perturbation $h_{\mu\nu}$. The linearized field equations
are then found to be
\be\label{finaleom}
\left(\square + m^2\right) G^{lin}_{\mu\nu} =0\, , \qquad R^{lin}  =0\, , 
\ee
where $G^{lin}_{\mu\nu}$ is the linearized Einstein tensor, and $R^{lin}$ the linearized Ricci scalar, which is zero by (\ref{trace}) since $K$ contains no  term linear in the metric perturbation.  The linearized Einstein tensor may be written in the form $[{\cal G}h]_{\mu\nu}$ where ${\cal G}$ is the following linear differential  operator, which we call the `Einstein' operator: 
\be\label{EO}
{\cal G}_{\mu\nu}{}^{\rho\sigma} = \frac{1}{2}
\varepsilon_{(\mu}{}^{\eta\rho}\varepsilon_{\nu)}{}^{\tau\sigma} \partial_\eta\partial_\tau\, . 
\ee
In momentum space, we may view ${\cal G}$ as a $6\times 6$ matrix. This matrix is not invertible but it maps the $3$-dimensional `physical' subspace of transverse metric perturbations to itself, and so defines a projected $3\times 3$ matrix ${\cal G}_\perp$. This projected matrix {\it is} invertible; it is for this reason that the $3D$ Einstein equation propagates no physical modes. The three eigenvectors of ${\cal G}_\perp$ are two traceless transverse metric perturbations describing massive modes of helicities $\pm2$, and the transverse trace, which describes a massive mode of zero helicity.  However, ${\cal G}_\perp$ is an operator of no definite sign; the eigenvalues of the helicity $\pm2$ modes are negative whereas the eigenvalue of the zero helicity mode is positive. This implies (given our conventions) that the helicity $\pm2$ modes are physical whereas the zero helicity mode would be  a `ghost' (i.e. have negative kinetic energy) were it not for the $R^{lin} =0$ constraint that removes precisely this mode. 

An implication of the foregoing is that linearized NMG  is equivalent to the Pauli-Fierz theory for a free massive spin 2 field. We will now demonstrate this equivalence directly. 
We begin with the observation  that (\ref{ouraction}) is equivalent to the action with Lagrangian density
\be\label{ouraction2}
{\cal L} = \frac{1}{\kappa^{2}}\sqrt{g}\left[R +   f^{\mu\nu}G_{\mu\nu} 
-\frac{1}{4}m^2\left(f^{\mu\nu}f_{\mu\nu} - f^2\right)\right] \, , 
 \ee
 where $f_{\mu\nu}$ is an auxiliary symmetric tensor field with trace $f= g^{\mu\nu}f_{\mu\nu}$.  Next, we expand  about a  Minkowski background, keeping only quadratic terms in the metric perturbation. The result is
\be
{\cal L}_2 =  \left(f^{\mu\nu}-\frac{1}{2}h^{\mu\nu}\right)\left[{\cal G} h\right]_{\mu\nu}
 - \frac{1}{4} m^2\left(f_{\mu\nu}f^{\mu\nu} -f^2\right)\, . 
\ee
Naturally, elimination of $f_{\mu\nu}$ yields the quadratic approximation to (\ref{ouraction}) about the Minkowski background solution, but we may instead eliminate $h_{\mu\nu}$; its field equation is ${\cal G} \left(h-f\right)=0$.  Because the Einstein operator is invertible on the space of transverse
symmetric tensors, the solution of this equation is $h_{\mu\nu}=f_{\mu\nu}$ up to 
a linearized gauge transformation which is irrelevant because the action is gauge invariant. 
Back-substitution now yields an action with Lagrangian density
\be
{\cal L} = \frac{1}{2}f^{\mu\nu} \left[{\cal G} f\right]_{\mu\nu}  - \frac{1}{4}m^2 \left(f^{\mu\nu}f_{\mu\nu} -f^2\right)\, .  
\ee
This is precisely the Pauli-Fierz theory for a massive spin 2 field $f_{\mu\nu}$.  The first term is the 
linearization of the EH term, which now has the  `right'  sign.

To further understand what is so special about the action (\ref{ouraction}), it is useful to consider it as a special case of the class of models in which $K$ is replaced by  $aK + bR^2$ for  constants $(a,b)$, not both zero. These models were investigated in \cite{Nishino:2006cb} for a `right-sign' EH term, in which case there are tachyons unless $a\le0$ and $b\ge0$, and only $a=0$ yields a ghost-free model, which propagates a single scalar mode.  Actually, it is well-known, at least for $4D$, that a Lagrangian density of the form ${\cal L}= \phi(R)$ is equivalent,  for some `suitable' class of functions $\phi$,  to GR coupled to a scalar field with a potential determined by the function  $\phi$; the history is summarized in \cite{Schmidt:2006jt}, where the extension to all $D\ge3$  is  also presented. For the `wrong-sign' EH term, there are tachyons unless $a\ge0$ and $b\le0$, and only $b=0$ yields a ghost-free model, which propagates
spin 2 modes of helicity $+2$ and $-2$. 

An analysis of the ${\cal N}=1,2$ supergravity extensions of the higher-derivative gravity with  $K\to aK+bR^2$ was also presented in  \cite{Nishino:2006cb}, again for `right-sign' EH term but the detailed results can be used to deduce some interesting consequences for `massive supergravities' which we define to be the supersymmetric extensions for $(a,b)=(1,0)$ with  `wrong-sign' EH term.  It will suffice to consider the bosonic fields.  For ${\cal N}=1$ there is a scalar  `auxiliary'  field  $S$ which actually is auxiliary for $b=0$. So the ${\cal N}=1$ massive supergravity is ghost-free and  propagates a  supermultiplet of spins $(2,3/2)$. For ${\cal N}=2$ there is a complex scalar auxiliary field which is, again, actually auxiliary only for $b=0$. There is also a real vector `auxiliary'  field;  remarkably, its action is the Proca          action for $b=0$. So the ${\cal N}=2$ massive supergravity is ghost-free and  propagates  a  supermultiplet of spins $(2,3/2,3/2,1)$ (each with two helicities). It is fairly clear that an ${\cal N}=4$ massive supergravity could be constructed similarly, using a `tensor calculus' derived from the ${\cal N}=2$ tensor calculus in $4D$, but  beyond that we can only speculate.

We now aim to  make contact with the parity-violating  topologically massive gravity.  We start from a `triple-master' action that depends on three second-rank tensor fields $(h,k,e)$ on $3D$ Minkowski spacetime. We assume that $h$ is a symmetric tensor but that $k$ and $e$ are {\it general} second-rank tensors. The Lagrangian density is 
\begin{eqnarray}\label{triple}
&&{\cal L}(h,k,e) = -\frac{1}{2\mu^2} \left(\mu h + 
2k\right)^{\mu\nu} \left[ {\cal G}(\mu h+ 2k)\right]_{\mu\nu} \nonumber \\
&&+\ \frac{1}{\mu }  \varepsilon^{\mu\nu\rho} \left(e + k\right)_\mu{}^\sigma \partial_\nu k_{\rho\sigma} -\frac{1}{4}\left( e^{\nu\mu}e_{\mu\nu} - e^2\right) \, , 
\end{eqnarray}
where $\mu$ is a mass parameter. Elimination of the auxiliary field $e$ yields
\be\label{lag(h,f)}
{\cal L}(h,k)=  - \frac{1}{2\mu} \left(\mu h + 4 k\right)_{\mu\nu} \left[{\cal G} h\right]^{\mu\nu}  + 
\frac{1}{\mu}  \varepsilon^{\mu\nu\rho} k_\mu{}^\alpha \partial_\nu k_{\rho\alpha}\, . 
\ee
The $k$-equation of motion has the solution 
\be 
k_\mu{}^\nu =  \frac{1}{2} \varepsilon^{\nu\sigma\lambda} \partial_\sigma h_{\lambda\mu} + \partial_{\mu} \xi^\nu\, , 
\ee
for arbitrary vector field $\xi$, which drops out on back-substitution; we thus get the Lagrangian density of linearized TMG: 
\be\label{lag(h)}
{\cal L}(h) = - \frac{1}{2}\left( h_{\mu\nu} + 
\frac{1}{\mu}\varepsilon_\nu{}^{\tau\sigma} \partial_\tau h_{\sigma\mu} \right)  \left[{\cal G} h\right]^{\mu\nu}\, . 
\ee
Note that the  EH term has the expected  `wrong' sign. Thus, the triple-master action is equivalent to linearized TMG,
but we now obtain two other equivalent actions as follows. 

Returning to (\ref{lag(h,f)}), we see that the $h$ equation of motion implies that $h_{\mu\nu} = -(2/\mu) k_{(\mu\nu)}$ modulo an irrelevant gauge transformation, and back-substitution then yields the Lagrangian density
\be\label{lag(f)}
{\cal L}(k) = \frac{2}{\mu^2} k^{\mu\nu}\left[{\cal G} k\right]_{\mu\nu}  + \frac{1}{\mu} \varepsilon^{\mu\nu\rho} k_\mu{}^\sigma\partial_\nu k_{\rho\sigma}\, . 
\ee
Note that the linearized EH term for $k$ has the `right' sign, and  that the second term depends on both the symmetric and antisymmetric parts of $k_{\mu\nu}$. This is  the `self-dual' model of  \cite{Deser:1990ay}.

Alternatively, we can return to (\ref{triple}) and eliminate $h$. Its equation of motion again implies that 
$h_{\mu\nu} = -(2/\mu) k_{(\mu\nu)}$ modulo an irrelevant gauge transformation, 
and back-substitution then gives
\be
{\cal L}(k,e) = \frac{1}{\mu} \varepsilon^{\mu\nu\rho} \left(k+e\right)_\mu{}^\sigma \partial_\nu k_{\rho\sigma} 
-\frac{1}{4} \left( e^{\nu\mu} e_{\mu\nu} -e^2\right) \, . 
\ee
The equation of motion for $k$ is 
\be
\varepsilon^{\mu\tau\rho}\partial_\tau \left(2 k_\rho{}^\nu + e_\rho{}^\nu\right) =0\, , 
\ee
which implies that $k_\mu{}^\nu = - \frac{1}{2} e_\mu{}^\nu + \partial_\mu\xi^\nu$ for arbitrary vector $\xi$ which, as before, drops out upon back-substitution to leave us with the Lagrangian density
\be\label{lag(e)}
{\cal L}(e) = - \frac{1}{4\mu} \varepsilon^{\mu\nu\rho} e_\mu{}^\sigma \partial_\nu e_{\rho\sigma}
-\frac{1}{4} \left(e^{\nu\mu}e_{\mu\nu} -e^2\right)\, . 
\ee
This is the `first-order' spin 2 model of \cite{Aragone:1986hm}. Its equations of motion 
can be shown to be equivalent to
\be\label{three}
\varepsilon^{\mu\tau\lambda} \partial_\tau e_\lambda{}^\nu + \mu\, e^{\nu\mu} =0\, , 
\quad \eta^{\mu\nu}e_{\mu\nu}=0\, ,  \quad   e_{[\mu\nu]} =0\, . 
\ee 
Iteration of the first-order differential equation, which implies that $\partial_\mu e^{\nu\mu}=0$,  yields the Klein-Gordon equation for $e_{\mu\nu}$, which is equivalent to the PF equations when combined with the algebraic constraints.
However, because the original equation was first-order, only one of the two spin 2 modes  of the PF theory is propagated. 

The equations of motion of the triple-master action imply that $e_{\mu\nu} = (2/\mu)[R^{lin}_{\mu\nu}-(1/4)\eta_{\mu\nu} R^{lin}]$, and using this in (\ref{three}) we arrive at the linearized TMG equations in the form
\be\label{linTMG}
{\cal O}_\mu{}^\rho(\mu) G^{lin}_{\rho\nu} =0 \, , \qquad R^{lin}=0\, , 
\ee
where ${\cal O}$ is the operator of the self-dual spin 1 theory \cite{Townsend:1983xs}:
\be
{\cal O} _\mu{}^\nu (\mu)  = \delta_\mu{}^\nu + \frac{1}{\mu} \varepsilon_\mu{}^{\tau\nu} \partial_\tau\, .  
\ee 
The tensor ${\cal O}G^{lin}$ is symmetric, despite appearances,  as a consequence of the linearized Bianchi identity. 
Let us now consider the alternative equations
\be\label{alternative}
\left[{\cal O}(-m_-) {\cal O}(m_+)\right]_\mu{}^\rho\,  G^{lin}_{\rho\nu} =0\, , \qquad R^{lin}=0\, . 
\ee
Evidently, these propagate helicities $\pm2$ with masses $m_\pm$,  so we recover (\ref{linTMG}) by taking  $m_-\to\infty$ for fixed $m_+=\mu$. If instead we set $m_+=m_-=m$ then we  get the parity-preserving equations (\ref{finaleom}).  In this sense, TMG is a  `square-root'  of the new massive gravity proposed here,  but both are actually special cases of a `general massive gravity' (GMG) theory with two mass parameters. To see this, we observe that  the equations  (\ref{alternative}) are equivalent to the linearization of the equation
\be\label{general}
G_{\mu\nu} + \frac{1}{\mu} C_{\mu\nu} + \frac{1}{2m^2} K_{\mu\nu} =0\, , 
\ee
where $C_{\mu\nu}$ is the Cotton tensor,
\be
C_{\mu\nu}=  (1/\sqrt{g})\varepsilon_\mu{}^{\tau\rho} D_\tau \left[R_{\rho\nu}- (1/4)g_{\rho\nu} R\right]\, , 
\ee
which arises from variation of a LCS term, and
\be
m^2 = m_+ m_- \, , \qquad \mu = m_+ m_-/(m_- -m_+)\, .
\ee
For $m\to\infty$ for fixed $\mu$ we recover TMG while $\mu\to\infty$ for fixed $m$ yields the  model defined by (\ref{ouraction}). 

We now turn to the cosmological extension of the GMG model  obtained by adding a cosmological term to the field equation (\ref{general}), as recently considered for TMG \cite{Li:2008dq,Carlip:2008jk}. 
Specifically, we consider the field equation
\be\label{generalcos}
 \lambda m^2 g_{\mu\nu} + \alpha\, G_{\mu\nu} + \frac{1}{\mu} C_{\mu\nu} + \frac{\beta}{2m^2}K_{\mu\nu} =0\, , 
\ee
where $\lambda$ is a dimensionless parameter, as are $\alpha$ and $\beta$, which we include for generality.  Let us seek maximally symmetric vacuum solutions for which 
\be\label{maxsym}
G_{\mu\nu} = \Lambda g_{\mu\nu}\, , 
\ee
for some `cosmological' constant $\Lambda$; for such solutions we have $C_{\mu\nu}=0$ and $K_{\mu\nu} = -\frac{1}{2} \Lambda^2 g_{\mu\nu}$. If $\beta=0$, but $\alpha\ne0$, the field equation is solved when $\Lambda = -(\lambda/\alpha) m^2$, which is the adS vacuum of `cosmological TMG'  for $\lambda/\alpha>0$. If $\beta\ne0$ then  the field equation is solved when 
\be\label{felam}
\beta\Lambda= 2 m^2 \left[ \alpha \pm \sqrt{\alpha^2+ \beta\lambda}\right] \qquad (\beta\ne0). 
\ee
There is no maximally symmetric vacuum unless $\beta\lambda \ge -\alpha^2$, and when this inequality is saturated there is a unique vacuum that is  de Sitter (dS) for $\beta/\alpha>0$ and anti-de Sitter (adS) for $\beta/\alpha<0$. Given $\alpha\ne0$, there are two inequivalent (a)dS vacua when $0>\beta\lambda> -\alpha^2$,  one of which becomes the Minkowski vacuum of our original massive gravity theory in the limit that $\lambda=0$. For $\beta\lambda>0$ there is one dS vacuum and one adS vacuum.  There will also be BTZ black holes \cite{Banados:1992wn} as these are locally isometric to adS, and it  would be interesting to see how their microscopic degrees of freedom  are  encoded in some holographically dual  $D=2$ field theory.  

In the context of ${\cal N}=1$ supergravity, a solution is supersymmetric if it admits a non-zero spinor field $\epsilon$ satisfying $(D_\mu + \frac{i}{2}S\gamma_\mu)\epsilon=0$, where $S$ is the auxiliary scalar, constant in a vacuum. The integrability condition is $G_{\mu\nu} =- S^2g_{\mu\nu}$, so $S^2=-\Lambda$ in a supersymmetric vacuum.  This condition is satisfied when $\beta=0$ and $\lambda/\alpha\ge0$, with $S= m\sqrt{\lambda/\alpha}$. For $\beta\ne0$ the $S$ field equation will be modified. Unfortunately, the modification  depends on unknown coefficients of $S^2R$ and $S^4$ terms in the action, so the status of adS vacua of cosmological super-GMG remains an interesting open question. 

Finally, in view of the ultra-violet finiteness of $4D$ gauge theories with ${\cal N}=4$ supersymmetry, it seems likely  that some  supersymmetric extension of the new massive 3D gravity presented here will be not just renormalizable but  ultra-violet finite.


\bigskip

\begin{acknowledgments}
PKT thanks the  EPSRC for financial support, and the University of Groningen for hospitality. 
This work is part of the research programme of the {\it Stichting voor Fundamenteel Onderzoek der Materie} (FOM). 
\end{acknowledgments}

\end{document}